\documentclass[10pt]{article}

\usepackage{fancyhdr}		           
\usepackage[dvips]{graphicx}	     
\usepackage{natbib}                
\usepackage{balance}               
\usepackage[margin=1in]{geometry}  
\usepackage{wrapfig}

\usepackage{booktabs}
\usepackage{threeparttable}
\usepackage[table]{xcolor}
\definecolor{Gray0}{gray}{0.75}
\definecolor{Gray1}{gray}{0.85}
\definecolor{Gray2}{gray}{0.95}

\usepackage[belowskip=3pt,aboveskip=0pt,font={small,it}]{caption}
\usepackage{proposal_ref}          
\usepackage{proposal_names}		     
\usepackage{proposal_figure}       
\usepackage{proposal_section}      

\usepackage[utopia]{mathdesign} 

\def\ion#1#2{#1\,{\sc #2}\relax}

\def\arcsec{\hbox{$^{\prime\prime}$}}

\newcommand{\mytitleL}{Understanding Alfv\'enic Waves in Flares}
\newcommand{\mytitleR}{}

\begin{document}

\begin{centering}
  \large
  \textbf{Science Objective: Understanding Energy Transport by Alfv\'enic Waves in Solar Flares} \\
  \vspace{2mm}  
  \small
  Jeffrey W. Reep$^{1,2}$, Harry P. Warren$^{2}$, James E. Leake$^{2}$, Lucas A. Tarr$^{1,2}$, Alexander J.B. Russell$^{3}$, Graham S. Kerr$^{4}$, Hugh S. Hudson$^{4,5}$ \\
  \vspace{1mm}
  \footnotesize
  $^1$National Research Council Postdoctoral Fellow, $^2$Naval Research Laboratory, $^3$University of Dundee, $^4$University of Glasgow, $^5$University of California, Berkeley  \\
\end{centering}

\section{Summary}

Solar flares are driven by the release of magnetic energy from reconnection events in the solar corona, whereafter energy is transported to the chromosphere, heating the plasma and causing the characteristic radiative losses.  In the collisional thick-target model (CTTM) \citep{brown1971,hudson1972}, electrons accelerated to energies exceeding 10\,keV traverse the corona and impact the chromosphere, where they deposit their energy through collisions with the much denser plasma in the lower atmosphere.  While there are undoubtedly high energy non-thermal electrons accelerated in flares, it is unclear whether these electron beams are the sole mechanism of energy transport, or whether they only dominate in certain phases of the flare's evolution.  Alfv\'enic waves are generated during the post-reconnection relaxation of magnetic field lines \citep{guidoni2010,longcope2012}, so it is important to examine their role in energy transport.

\citet{emslie1982} proposed that Alfv\'enic waves, produced in the corona and traveling downwards in the solar atmosphere, can sufficiently explain observed heating at the temperature minimum region in large flares.  More recently, \citet{fletcher2008} proposed that these waves were able to transport the released magnetic energy to heat the entire chromosphere, and accelerate electrons {\it in situ} in the chromosphere, thereby producing the characteristic hard X-ray (HXR) emission of flares.  Using the formalism of wave dissipation developed by \citet{emslie1982} and \citet{russell2013}, \citet{reep2016} showed that high frequency Alfv\'enic waves can heat the upper chromosphere, and produce an atmospheric response that is strikingly similar to the CTTM.  This result was confirmed by \citet{kerr2016}, who concluded that despite the superficial similarities, the two mechanisms could be distinguished in chromospheric emission lines such as the \ion{Mg}{ii} k line.  These predictions have never been tested directly with observations.

Both \citet{reep2016} and \citet{kerr2016} found that wave frequencies $\gtrsim 1$\,Hz strongly heat the upper chromosphere.  It is not known at present what Alfv\'enic wave frequencies might occur in solar flares, although mHz waves are seen ubiquitously in active regions \citep{tomczyk2007,mcintosh2011}.  Waves with frequencies as high as $0.1$\,Hz were observed with TRACE \citep{deforest2004}, while $1$\,Hz waves have been seen in coronal lines during total eclipses \citep{pasachoff2002}.  Models of magnetohydrodynamic waves in arcades do predict the presence of Alfv\'en waves with frequencies  higher than $0.1$\,Hz (e.g. \citealt{oliver1993}), although they have not yet been observationally confirmed.  In order to resolve frequencies $\gtrsim 1$\,Hz, an imager with cadence of $\lesssim 1$\,s is required.  

An important consideration is non-thermal line broadening, commonly observed in flares, which may be attributable to unresolved wave motions.  For example, \citet{milligan2011} measured line widths in  two \ion{Fe}{XIV} lines during a small flare with Hinode-EIS, and found a large non-thermal component ($\approx 50$\,km~s$^{-1}$) that could not be explained by opacity or pressure broadening.  \citet{polito2015} similarly found large non-thermal velocities in strongly blue-shifted \ion{Fe}{XXI} emission measured with IRIS.  The role of wave motions in producing non-thermal line broadening could be ascertained with spectra at a cadence high enough to resolve these wave motions.


\section{Proposed Instrumentation}

A spectrometer with a cadence of less than 1\,s and spectral coverage of the transition region (TR) and chromosphere could measure this energy transport.  Specifically, by measuring spectral lines from ions of many ionization stages and formation depths, the transport of energy could be followed through the impulsive phase of a solar flare, strongly constraining our models of energy release.  In order to resolve the waves, high cadence imaging is similarly required.  


In this white paper, we therefore propose that these questions could be examined with an improved slit spectrograph with an explicit aim towards extremely high cadence spectra in the extreme ultraviolet range, which can only be seen from space.  The instrument should be designed to (1) measure the depths of energy deposition, (2) measure the travel times between these depths, (3) resolve high frequency waves, and (4) discriminate the atmospheric response between electron beams and Alfv\'en waves.  A slit-jaw imager, whereby light surrounding the slit is reflected onto broadband filters (e.g. IRIS, \citealt{depontieu2014}), with extremely high cadence of $\approx 0.1$\,s could suitably resolve high frequency wave motions.  Finally, to measure information about the non-thermal electron population, a hard X-ray spectrometer (e.g. MinXSS, \citealt{mason2016}), which need not have spatial resolution, could supplement these observations to determine the presence of non-thermal electrons via bremsstrahlung signatures and estimate their energy distribution, as well as giving estimates of the flaring loop temperatures from thermal continuum emissions.

The magnetic field strength and structure determines how waves will propagate through the plasma, and to determine this we plan to rely on coordination with the current and next generation of ground-based solar telescopes, specifically the Daniel K. Inouye Solar Telescope (DKIST, \citealt{elmore2014}, expected first light 2019) and European Solar Telescope (EST, \citealt{collados2013}, expected first light 2026), both currently under construction.  The instrument suites at these facilities will complement our proposed observations with very high spatial and temporal resolution measurements of chromospheric lines and magnetic field structure.  The current generation of ground--based instrumentation already makes routine polarimetric observations of photospheric and chromospheric lines (e.g. \ion{Fe}{I} 6302\,\AA{} and \ion{Ca}{II} 8542\,\AA{}, \citealt{keller2003}) from which the magnetic field is determined by inversion using an ever-expanding set of techniques (e.g. \citealt{sanka2006}).

DKIST will break from previous ground-based solar telescopes by implementing a queue system for observations which will allow for increased target--of--opportunity observations to be collected, e.g., during flares.  Temporal cadence of observations is limited by the camera speed, but will surpass what we propose for the satellite observations.  For instance, DLNIRSP camera speed is $\sim 33$\,ms, and the small FOV ($5\arcsec\times5\arcsec$) may be tessellated, with each step requiring an addition $~50$\,ms, so that a $25\arcsec\times25\arcsec$ FOV may be generated in as little as $2$\,s.  This is at the highest resolution, which is not always necessary.  Furthermore, the spectroscopic instruments of DKIST---DLNIRSP and ViSP---are designed to be used simultaneously, and to each individually measure several spectral windows simultaneously.  Their addition of visible and infrared wavelengths at very high spatial, temporal, and spectral resolution will compliment our proposed EUV mission to determine where, and by what mechanism, energy is deposited during flares.

\begin{figure}[t]
  \centerline{\includegraphics[clip,width=\linewidth]{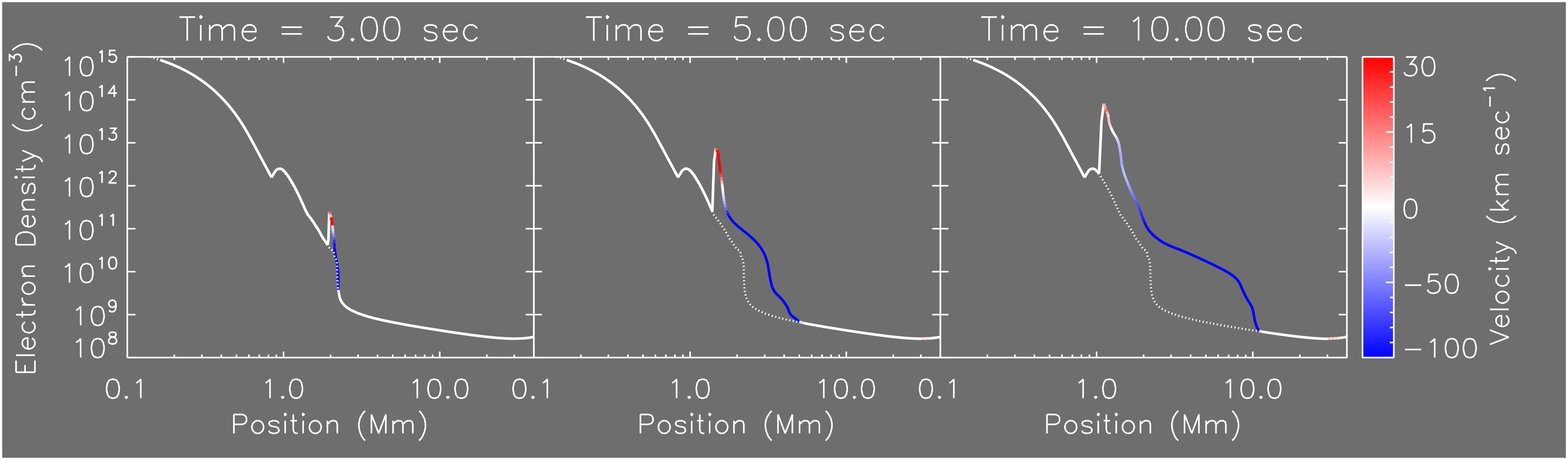}}
    \centerline{\includegraphics[clip,width=\linewidth]{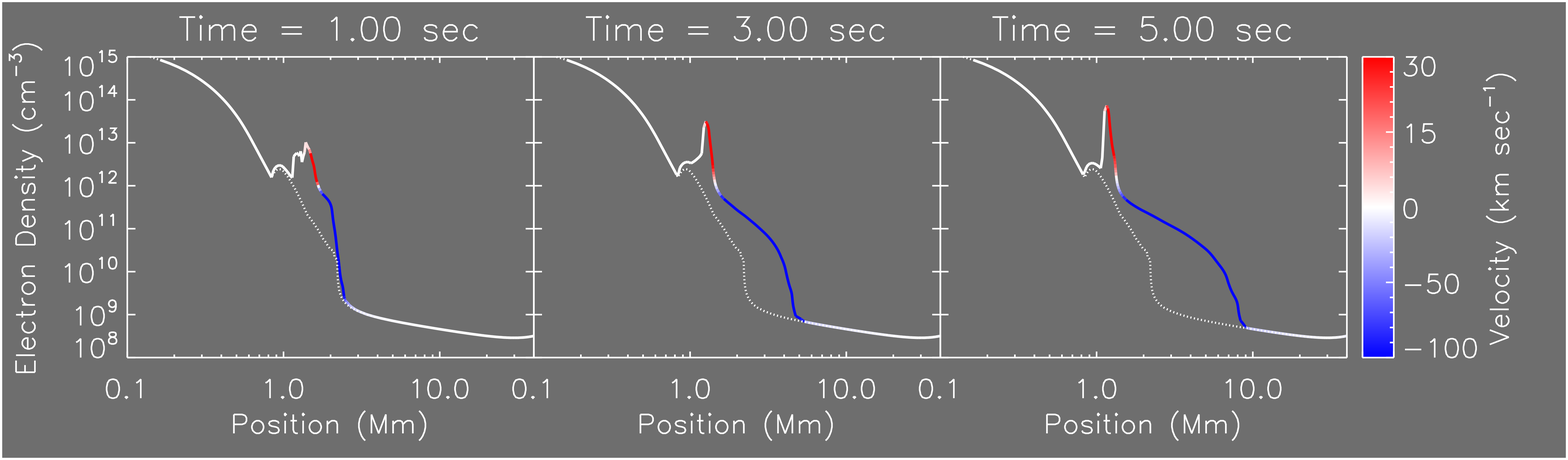}}
  \vspace{2mm}
  \caption{The electron density along a flare loop in a simulation of Alfv\'enic wave heating (top) and electron beam heating (bottom), at three select times.  As the wave pulse propagates through the chromosphere dissipating its energy, the local temperature rises, causing an increase in the ionization of hydrogen, which in turn increases the electron density at ever-increasing depths.  Contrast this with the beam case, where the increase in electron density occurs at approximately the same depth during the early impulsive phase.  The local flow-speed velocity is colored blue where material is up-flowing and red where it's down-flowing.  The dotted lines indicate the initial density.  The x-axis is displayed on a logarithmic scale to emphasize the chromosphere.}
  \label{fig:density}
\end{figure}

\section{Simulations}

We briefly present numerical simulations performed with the HYDRAD code \citep{bradshaw2003,bradshaw2013}, adapted to include heating by a beam of electrons or heating due to the dissipation of Alfv\'enic waves.  A $10$\,keV electron travels at $\approx60$\,Mm~s$^{-1}$, so we neglect travel times for electron beams in general.  However, waves travel at the Alfv\'en speed, which is often $5-10$\,Mm~s$^{-1}$ in the corona, though considerably less in the chromosphere, so time delays in wave heating cannot be neglected.  The simulation presented here has modified the code used in \citet{reep2016} to therefore trace the location of the waves with time.

Figure \ref{fig:density} shows the electron density evolution in an example simulation of heating by Alfv\'enic wave dissipation (top plots).  The wave pulse, injected at the apex of the loop, propagates downwards at the Alfv\'en speed ($\approx 9$\,Mm~s$^{-1}$ in the corona), reaching the chromosphere in approximately 3 seconds.  The local increase in density drives an increased dissipation of the wave via ion-neutral and electron-ion collisions, which in turn heats the ambient plasma.  The heat deposition raises the temperature, which in turn raises the electron density as the hydrogen ionization fraction increases, and drives evaporation and condensation flows due to the local pressure expansion.  The wave pulse continues to propagate downwards, similarly dissipating its Poynting flux at ever greater depths.  These plots can be contrasted with an electron beam heating simulation (bottom plots).  The electrons reach the chromosphere essentially instantly, where they are stopped due to collisions with the ambient plasma.  The depth at which this occurs is approximately constant, and it does not propagate to deeper depths, unlike the wave heating case.

\begin{figure}
\includegraphics[clip,width=0.5\linewidth]{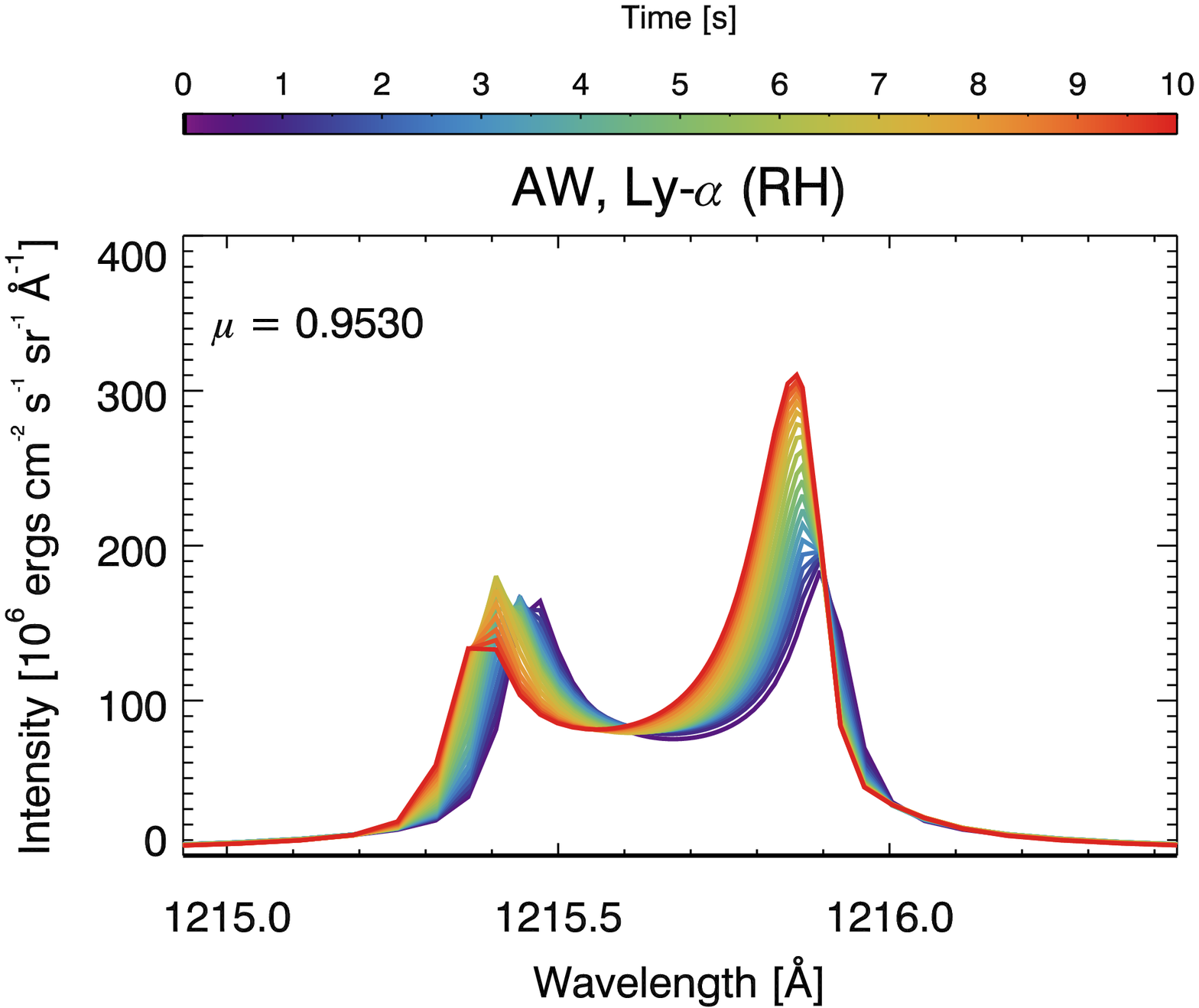}
\includegraphics[clip,width=0.5\linewidth]{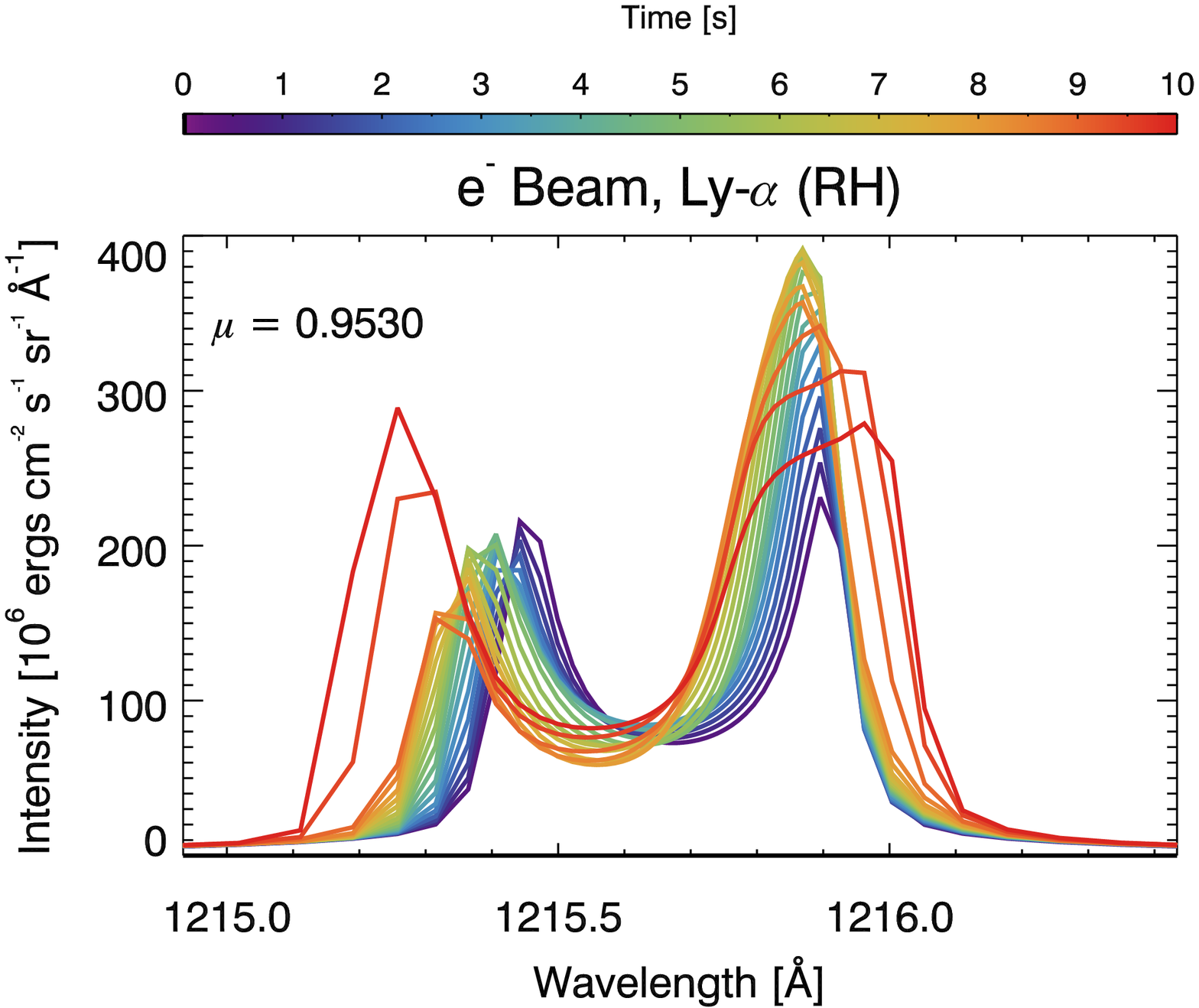}
\caption{Lyman-$\alpha$ calculated at 0.5-second cadence from the wave (left) and beam heating (right) simulations presented in \citet{kerr2016}.  The line evolves quickly and the two heating mechanisms show divergent properties in terms of intensities, widths, blue-to-red wing ratios, and Doppler shifts.}
\label{fig:lyalpha}
\end{figure}

The depth at which heat is deposited therefore can discriminate between these two mechanisms.  To measure this precisely, observations of chromospheric lines that form at varying depths at high cadence are required.  \citet{kerr2016}, for example, showed that \ion{Mg}{II} and \ion{Ca}{II} respond differently to wave heating and electron beam heating in the mid chromosphere.  We add the example of Lyman-$\alpha$ to this, shown in Figure \ref{fig:lyalpha}, calculated from the same simulations and with the same methodology of \citet{kerr2016}.  Lyman-$\alpha$ forms primarily at the top of the chromosphere, with the Lyman continuum formed at slightly lower heights \citep{vernazza1981,carlsson2002}.  The RAISE sounding rocket \citep{laurent2016} has recently demonstrated the viability of observing this line at a cadence $> 5$\,Hz.

Therefore, in order to (1) measure what frequency waves occur in flares, (2) trace the location of heating in the chromosphere, and (3) distinguish heating mechanisms, a high cadence spectrograph with a bandpass centered on important chromospheric lines with imaging capability via a slit-jaw imager would be the ideal instrument.  The following lines offer good chromospheric and TR coverage and are intense enough to observe at high cadence: Lyman-$\alpha$, Lyman-$\beta$, \ion{He}{I} $584.3$\,\AA, \ion{He}{II} $303.8$\,\AA, \ion{Mg}{II} h and k, \ion{Si}{II} $1194.5$\,\AA\ and $1264.7$\,\AA.  In combination with the optical lines measured with DKIST and EST, the motion of magnetohydrodynamic waves through the chromosphere can be traced so that the role of waves in flares will be understood. 


\clearpage

This work was performed while JWR held an NRC Research Associateship award at the US Naval Research Laboratory with the support of NASA.  LAT was supported by the Chief of Naval Research.


\begin{thebibliography}{27}
\expandafter\ifx\csname natexlab\endcsname\relax\def\natexlab#1{#1}\fi

\bibitem[{{\it {Bradshaw} and {Cargill}\/}(2013)}]{bradshaw2013}
{Bradshaw}, S.~J., and P.~J. {Cargill}, {The Influence of Numerical Resolution
  on Coronal Density in Hydrodynamic Models of Impulsive Heating}, {\it
  \apj\/}, {\it 770\/}, 12, 2013.

\bibitem[{{\it {Bradshaw} and {Mason}\/}(2003)}]{bradshaw2003}
{Bradshaw}, S.~J., and H.~E. {Mason}, {A self-consistent treatment of radiation
  in coronal loop modelling}, {\it \aap\/}, {\it 401\/}, 699--709, 2003.

\bibitem[{{\it {Brown}\/}(1971)}]{brown1971}
{Brown}, J.~C., {The Deduction of Energy Spectra of Non-Thermal Electrons in
  Flares from the Observed Dynamic Spectra of Hard X-Ray Bursts}, {\it
  \solphys\/}, {\it 18\/}, 489--502, 1971.

\bibitem[{{\it {Carlsson} and {Stein}\/}(2002)}]{carlsson2002}
{Carlsson}, M., and R.~F. {Stein}, {Dynamic Hydrogen Ionization}, {\it \apj\/},
  {\it 572\/}, 626--635, 2002.

\bibitem[{{\it {Collados} et~al.\/}(2013){\it {Collados},
  et~al.\/}}]{collados2013}
{Collados}, M., et~al., {The European Solar Telescope}, {\it Memorie della
  Societa Astronomica Italiana\/}, {\it 84\/}, 379, 2013.

\bibitem[{{\it {De Pontieu} et~al.\/}(2014){\it {De Pontieu},
  et~al.\/}}]{depontieu2014}
{De Pontieu}, B., et~al., {The Interface Region Imaging Spectrograph (IRIS)},
  {\it \solphys\/}, {\it 289\/}, 2733--2779, 2014.

\bibitem[{{\it {DeForest}\/}(2004)}]{deforest2004}
{DeForest}, C.~E., {High-Frequency Waves Detected in the Solar Atmosphere},
  {\it \apjl\/}, {\it 617\/}, L89--L92, 2004.

\bibitem[{{\it {Elmore} et~al.\/}(2014){\it {Elmore}, et~al.\/}}]{elmore2014}
{Elmore}, D.~F., et~al., {The Daniel K. Inouye Solar Telescope first light
  instruments and critical science plan}, in {\it Ground-based and Airborne
  Instrumentation for Astronomy V\/}, vol. 9147 of {\it SPIE\/}, p. 914707,
  2014.

\bibitem[{{\it {Emslie} and {Sturrock}\/}(1982)}]{emslie1982}
{Emslie}, A.~G., and P.~A. {Sturrock}, {Temperature minimum heating in solar
  flares by resistive dissipation of Alfven waves}, {\it \solphys\/}, {\it
  80\/}, 99--112, 1982.

\bibitem[{{\it {Fletcher} and {Hudson}\/}(2008)}]{fletcher2008}
{Fletcher}, L., and H.~S. {Hudson}, {Impulsive Phase Flare Energy Transport by
  Large-Scale Alfv{\'e}n Waves and the Electron Acceleration Problem}, {\it
  \apj\/}, {\it 675\/}, 1645--1655, 2008.

\bibitem[{{\it {Guidoni} and {Longcope}\/}(2010)}]{guidoni2010}
{Guidoni}, S.~E., and D.~W. {Longcope}, {Shocks and Thermal Conduction Fronts
  in Retracting Reconnected Flux Tubes}, {\it \apj\/}, {\it 718\/}, 1476--1490,
  2010.

\bibitem[{{\it {Hudson}\/}(1972)}]{hudson1972}
{Hudson}, H.~S., {Thick-Target Processes and White-Light Flares}, {\it
  \solphys\/}, {\it 24\/}, 414--428, 1972.

\bibitem[{{\it {Keller} et~al.\/}(2003){\it {Keller}, {Harvey}, and {Solis
  Team}\/}}]{keller2003}
{Keller}, C.~U., J.~W. {Harvey}, and {Solis Team}, {The SOLIS
  Vector-Spectromagnetograph}, in {\it Solar Polarization\/}, edited by
  J.~{Trujillo-Bueno} and J.~{Sanchez Almeida}, vol. 307 of {\it Astronomical
  Society of the Pacific Conference Series\/}, p.~13, 2003.

\bibitem[{{\it {Kerr} et~al.\/}(2016){\it {Kerr}, {Fletcher}, {Russell}, and
  {Allred}\/}}]{kerr2016}
{Kerr}, G.~S., L.~{Fletcher}, A.~J.~B. {Russell}, and J.~C. {Allred},
  {Simulations of the Mg II k and Ca II 8542 lines from an Alfv{\'E}n
  Wave-heated Flare Chromosphere}, {\it \apj\/}, {\it 827\/}, 101, 2016.

\bibitem[{{\it {Laurent} et~al.\/}(2016){\it {Laurent},
  et~al.\/}}]{laurent2016}
{Laurent}, G.~T., et~al., {The Rapid Acquisition Imaging Spectrograph
  Experiment (RAISE) Sounding Rocket Investigation}, {\it Journal of
  Astronomical Instrumentation\/}, {\it 5\/}, 1640,006--34, 2016.

\bibitem[{{\it {Longcope} and {Tarr}\/}(2012)}]{longcope2012}
{Longcope}, D.~W., and L.~{Tarr}, {The Role of Fast Magnetosonic Waves in the
  Release and Conversion via Reconnection of Energy Stored by a Current Sheet},
  {\it \apj\/}, {\it 756\/}, 192, 2012.

\bibitem[{{\it {Mason} et~al.\/}(2016){\it {Mason}, et~al.\/}}]{mason2016}
{Mason}, J.~P., et~al., {Miniature X-Ray Solar Spectrometer: A
  Science-Oriented, University 3U CubeSat}, {\it Journal of Spacecraft and
  Rockets\/}, {\it 53\/}, 328--339, 2016.

\bibitem[{{\it {McIntosh} et~al.\/}(2011){\it {McIntosh}, {de Pontieu},
  {Carlsson}, {Hansteen}, {Boerner}, and {Goossens}\/}}]{mcintosh2011}
{McIntosh}, S.~W., B.~{de Pontieu}, M.~{Carlsson}, V.~{Hansteen}, P.~{Boerner},
  and M.~{Goossens}, {Alfv{\'e}nic waves with sufficient energy to power the
  quiet solar corona and fast solar wind}, {\it Nature\/}, {\it 475\/},
  477--480, 2011.

\bibitem[{{\it {Milligan}\/}(2011)}]{milligan2011}
{Milligan}, R.~O., {Spatially Resolved Nonthermal Line Broadening during the
  Impulsive Phase of a Solar Flare}, {\it \apj\/}, {\it 740\/}, 70, 2011.

\bibitem[{{\it {Oliver} et~al.\/}(1993){\it {Oliver}, {Ballester}, {Hood}, and
  {Priest}\/}}]{oliver1993}
{Oliver}, R., J.~L. {Ballester}, A.~W. {Hood}, and E.~R. {Priest},
  {Magnetohydrodynamic waves in a potential coronal arcade}, {\it \aap\/}, {\it
  273\/}, 647, 1993.

\bibitem[{{\it {Pasachoff} et~al.\/}(2002){\it {Pasachoff}, {Babcock},
  {Russell}, and {Seaton}\/}}]{pasachoff2002}
{Pasachoff}, J.~M., B.~A. {Babcock}, K.~D. {Russell}, and D.~B. {Seaton},
  {Short-Period Waves That Heat the Corona Detected at the 1999 Eclipse}, {\it
  \solphys\/}, {\it 207\/}, 241--257, 2002.

\bibitem[{{\it {Polito} et~al.\/}(2015){\it {Polito}, {Reeves}, {Del Zanna},
  {Golub}, and {Mason}\/}}]{polito2015}
{Polito}, V., K.~K. {Reeves}, G.~{Del Zanna}, L.~{Golub}, and H.~E. {Mason},
  {Joint High Temperature Observation of a Small C6.5 Solar Flare With
  Iris/Eis/Aia}, {\it \apj\/}, {\it 803\/}, 84, 2015.

\bibitem[{{\it {Reep} and {Russell}\/}(2016)}]{reep2016}
{Reep}, J.~W., and A.~J.~B. {Russell}, {Alfv\'enic Wave Heating of the Upper
  Chromosphere in Flares}, {\it \apjl\/}, {\it 818\/}, L20, 2016.

\bibitem[{{\it {Russell} and {Fletcher}\/}(2013)}]{russell2013}
{Russell}, A.~J.~B., and L.~{Fletcher}, {Propagation of Alfv{\'e}nic Waves from
  Corona to Chromosphere and Consequences for Solar Flares}, {\it \apj\/}, {\it
  765\/}, 81, 2013.

\bibitem[{{\it {Sankarasubramanian} et~al.\/}(2006){\it {Sankarasubramanian},
  et~al.\/}}]{sanka2006}
{Sankarasubramanian}, K., et~al., {The Diffraction Limited
  Spectro-Polarimeter}, in {\it Astronomical Society of the Pacific Conference
  Series\/}, edited by R.~{Casini} and B.~W. {Lites}, vol. 358 of {\it
  Astronomical Society of the Pacific Conference Series\/}, p. 201, 2006.

\bibitem[{{\it {Tomczyk} et~al.\/}(2007){\it {Tomczyk}, {McIntosh}, {Keil},
  {Judge}, {Schad}, {Seeley}, and {Edmondson}\/}}]{tomczyk2007}
{Tomczyk}, S., S.~W. {McIntosh}, S.~L. {Keil}, P.~G. {Judge}, T.~{Schad}, D.~H.
  {Seeley}, and J.~{Edmondson}, {Alfv{\'e}n Waves in the Solar Corona}, {\it
  Science\/}, {\it 317\/}, 1192, 2007.

\bibitem[{{\it {Vernazza} et~al.\/}(1981){\it {Vernazza}, {Avrett}, and
  {Loeser}\/}}]{vernazza1981}
{Vernazza}, J.~E., E.~H. {Avrett}, and R.~{Loeser}, {Structure of the solar
  chromosphere. III - Models of the EUV brightness components of the
  quiet-sun}, {\it \apjs\/}, {\it 45\/}, 635--725, 1981.

\end{thebibliography}

\end{document}